# Modelling Interdependencies between the Electricity and Information Infrastructures

Jean-Claude Laprie, Karama Kanoun, Mohamed Kaâniche

LAAS-CNRS, Université de Toulouse, France
{laprie, kanoun, kaaniche@laas.fr}

**Abstract.** The aim of this paper is to provide qualitative models characterizing interdependencies related failures of two critical infrastructures: the electricity infrastructure and the associated information infrastructure. The interdependencies of these two infrastructures are increasing due to a growing connection of the power grid networks to the global information infrastructure, as a consequence of market deregulation and opening. These interdependencies increase the risk of failures. We focus on cascading, escalating and common-cause failures, which correspond to the main causes of failures due to interdependencies. We address failures in the electricity infrastructure, in combination with accidental failures in the information infrastructure, then we show briefly how malicious attacks in the information infrastructure can be addressed.

## 1 Introduction

In the past decades, the electric power grid experienced several severe failures that affected the power supply to millions of customers. The most recent one occurred in November 2006 in Western Europe when a shutdown of a high-voltage line in Germany resulted in massive power failures in France and Italy as well as in parts of Spain, Portugal, the Netherlands, Belgium and Austria, and even extended as far as Morocco. About ten million customers were affected by this failure. Similar major blackouts with even more severe consequences have occurred in summer 2003 in the United States, in Canada and in Italy [1, 2]. These events highlight the vulnerability of the electric grid infrastructures and their interdependencies. The large geographic extension of power failures effects is related to i) the high interconnectivity of power grid transmission and distribution infrastructures and to ii) the multiple interdependencies existing between these infrastructures and the information infrastructures supporting the control, the monitoring, the maintenance and the exploitation of power supply systems. An *interdependency* is a bidirectional relationship between two infrastructures through which the state of each infrastructure influences or is correlated to the state of the other. Clearly there is a need to analyze and model critical infrastructures in the presence of interdependencies in order to understand i) how such interdependencies may contribute to the occurrence of large outages and ii) how to reduce their impact.

This paper focuses on two interdependent infrastructures: the electric power infrastructure and the information infrastructures supporting management, control and maintenance functionality. More specifically, it addresses modelling and analysis of interdependency-related failures between these infrastructures.

We concentrate on cascading, escalating and common-cause failures, which correspond to the main causes of interdependency-related failures. We model the infrastructures globally, not explicitly modelling their components. The models presented are qualitative ones. They describe scenarios that are likely to take place when failures occur. The models are built based on assumptions related to the behaviour of the infrastructures as resulting from their mutual interdependencies.

In the remainder of this paper, we will first address failures in the electricity infrastructure and accidental failures in the information infrastructure, considering the three classes of interdependencies, then we will illustrate briefly how malicious attacks of information infrastructures can be addressed.

Section 2 presents the background and related work. Sections 3 and 4 are dedicated to modelling interdependencies taking into account failures in the information infrastructure and in the electricity infrastructure. Section 3 addresses accidental failures in the information infrastructure while Section 4 addresses malicious attacks. Section 5 concludes the paper.

## 2. Background and Related Work

Interdependencies increase the vulnerability of the corresponding infrastructures as they give rise to multiple error propagation channels from one infrastructure to another that make them more prone to exposure to accidental as well as to malicious threats. Consequently the impact of infrastructure components failures and their severity can be exacerbated and are generally much higher and more difficult to foresee, compared to failures confined to single infrastructures. As an example, most major power grid blackouts that have occurred in the past were initiated by a single event (or multiple related events such as a power grid equipment failure that is not properly handled by the SCADA, i.e., Supervisory Control And Data Acquisition, system) that gradually leads to cascading failures and eventual collapse of the entire system [2].

Infrastructure interdependencies can be categorized according to various dimensions in order to facilitate their identification, understanding and analysis. Six dimensions have been identified in [3]. They correspond to: a) the type of interdependencies (physical, cyber, geographic, and logical), b) the infrastructure environment (technical, business, political, legal, etc.), c) the couplings among the infrastructures and their effects on their response behaviour (loose or tight, inflexible or adaptive), d) the infrastructure characteristics (organisational, operational, temporal, spatial), e) the state of operation (normal, stressed, emergency, repair), the degree to which the infrastructures are coupled, f) the type of failure affecting the infrastructures (common-cause, cascading, escalating). Other classifications have also been proposed in [4-6, 29]. In particular, the study reported in [29], based on 12 years public domain failure data, provides useful insights about the sources of failures affecting critical infrastructures, their propagation and their impact on public life, considering in particular the interdependencies between the communication and information technology infrastructure and other critical infrastructures such as electricity, transportation, financial services, etc.

Referring to the classification of [3], our work addresses the three types of failures that are of particular interest when analyzing interdependent infrastructures: i) cascading failures, ii) escalating failures, and iii) common cause failures. Definitions are as follows:

- *Cascading failures* occur when a failure in one infrastructure causes the failure of one or more component(s) in a second infrastructure.
- *Escalating failures* occur when an existing failure in one infrastructure exacerbates an independent failure in another infrastructure, increasing its severity or the time for recovery and restoration from this failure.
- *Common cause failures* occur when two or more infrastructures are affected simultaneously because of some common cause.

It is noteworthy that these classes of failures are not independent; e.g., common-cause failures can cause cascading failures [7].

Among the three relevant types of failures in interdependent infrastructures, the modelling of cascading failures has received increasing interest in the past years, in particular after the large blackouts of electric power transmission systems in 1996 and 2003. Several research papers and modelling studies have been published on this topic in particular by the Consortium for Electric Reliability Technology Solutions (CERTS) in the United-States [8]. A large literature has been dedicated recently to the elaboration of analytic or simulation based models that are able to capture the dynamics of cascading failures and blackouts. A brief review of related research addressing this topic is given hereafter. A more detailed state-of-the art can be found in [9, 10].

In [11, 12], the authors present an idealised probabilistic model of cascading failures called CASCADE that is simple enough to be analytically tractable. It describes a general cascading process in which component failures weaken and further load the system so that components failures are more likely. This model describes a finite number of identical components that fail when their loads exceed a threshold. As components fail, the system becomes more loaded, since an amount of load is transferred to the other components, and cascading failures of further components become likely. This cascade model and variants of it have been approximated in [13-15] by a Galton-Watson branching process in which failures occur in stages, with each failure giving rise to a Poisson distribution of failures at the next stage.

The models mentioned above do not take into account the characteristics of power systems. An example of a cascading failures model for a power transmission system is discussed in [16]. The proposed model represents transmission lines, loads, generators and the operating limits on these components. Blackout cascades are essentially instantaneous events due to dynamical redistribution of power flows and are triggered by probabilistic failures of overloaded lines. In [17] a simulation model is proposed to calculate the expected cost of failures, taking into account time-dependent phenomena such a cascade tripping of elements due to overloads, malfunction of the protection system, potential power system instabilities and weather conditions. Other examples emphasizing different aspects of the problem have been proposed e.g., in [18-21], in which hidden failures of the protection system are represented. Their approach uses a probabilistic model to simulate the incorrect tripping of lines and generators due to hidden failures of line or generator protection systems. The distribution of power system blackout size is obtained using importance sampling and Monte-Carlo simulation.

Recently, new approaches using complex networks theory have been also proposed for modelling cascading failures [22-27]. These models are based on the analysis of the topology of the network characterizing the system and the evaluation of the resilience of the network to the removal of nodes and arcs, due either to random failures or to malicious attacks).

All the models discussed above adopt a simplified representation of the power system, assuming that the overloading of system components eventually leads to the collapse of the global system. However, these models do not take into account explicitly the complex interactions and interdependencies between the power infrastructure and the ICT infrastructures. Moreover, the modelling of escalating failures is not addressed. Further work is needed in these directions. This paper presents a preliminary attempt at filling such gaps.

## 3. Accidental failures in the information infrastructure

Our aim is to model the infrastructures behaviour together, when taking into account the impact of accidental failures in the information infrastructure and failures in the electricity infrastructure, as well as their effects on both infrastructures. Modelling is carried out progressively:

- First, we model cascading failures by analysing the constraints one infrastructure puts on the other one, assuming that the latter was in a working state when an event occurs in the other one.
- Then, we address cascading and escalating failures considering successively:
  - constraints of the information infrastructure on the electricity infrastructure,
  - constraints both ways (of the information infrastructure on the electricity infrastructure and of the electricity infrastructure on the information infrastructure).
- Finally, we address common-cause failures.

For the sake of clarity, and in order to avoid any confusion between the two infrastructures, we use specialized but similar terms for the two infrastructures states and events as indicated by Table 1.

**Table 1.** States and events of the infrastructures

| Information Infrastructure | Electricity Infrastructure |
|---|---|
| i-failure | e-failure |
| i-restoration | e-restoration |
| i-working | e-working |
| Partial i-outage | Partial e-outage, e-lost |
| i-weakened | e-weakened |

### 3.1. Modelling cascading failures

We first analyse the impact of accidental i-failures on each infrastructure assuming that the electricity infrastructure is in an e-working state, then we analyse the impact of e-failures on each infrastructure assuming that the and information infrastructure is in an i-working state, before considering the combined impact of i- and e-failures in Section 3.2.

**3.1.1 Impact of information infrastructure failures (i-failures).** Accidental i-failures, hardware- or software-induced, affecting the information infrastructure can be:

- Masked (unsignalled) i-failures, leading to latent errors.
- Signalled i-failures.

Latent errors can be:

- Passive (i.e., without any action on the electricity infrastructure), but keeping the operators uninformed of possible disruptions occurring in the electricity infrastructure.
- Active, provoking undue configuration changes in the electricity infrastructure.

After signalled i-failures, the information infrastructure is in a partial i-outage state: the variety of functions and components of the information infrastructure, and its essential character of large network make unlikely total outage. Latent errors can accumulate. Signalled i-failures may take place when the information infrastructure is in latent error states. When the information infrastructure is in a partial i-outage state, i-restoration is necessary to bring it back to an i-working state.

Fig. 1-a gives the state machine model of the information infrastructure taking into account its own failures. It is noteworthy that all states are presented by several boxes, meaning that a state corresponds in reality to a group of different states that are considered as equivalent with respect to the classification given in Table 1. For example all states with only one busbar isolated can be considered as equivalent irrespective of which busbar is isolated.

We assume that an i-failure puts some constraints on the electricity infrastructure (i.e., cascading failure), leading to a weakened electricity infrastructure (e.g., with a lower performance, unduly isolations, or unnecessary off-line trips of production plants or of transmission lines).

From an e-weakened state, a configuration restoration leads the electricity infrastructure back into a working state, because no e-failures occurred in the electricity infrastructure. Accumulation of untimely configuration changes, may lead to e-lost state (i.e., a blackout state), from which an e-restoration is required to bring back the electricity infrastructure into an e-working state. Fig. 1-b shows the constraint that the information infrastructure puts on the electricity infrastructure when the latter is in an e-working state.

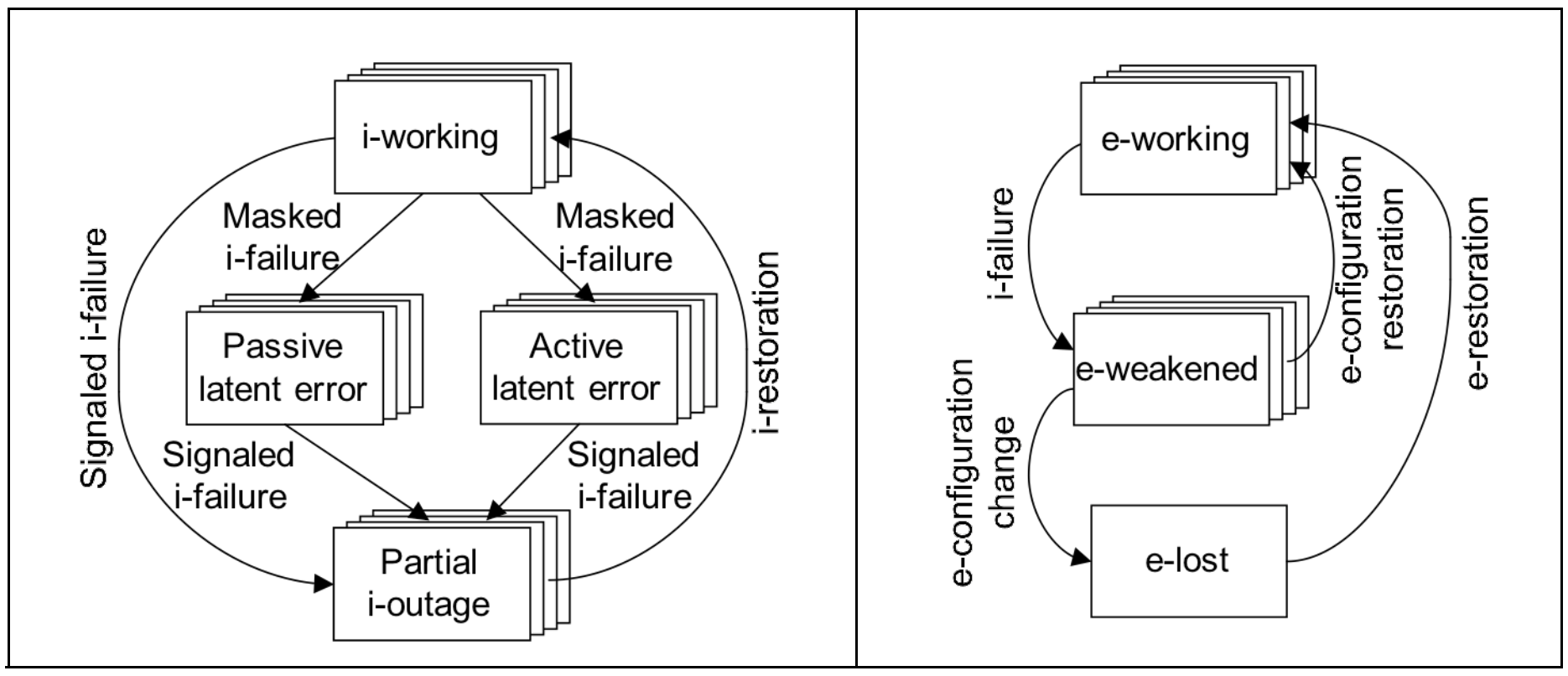

-a- Failure behaviour of the information infrastructure

-b- Impact of the information infrastructure on the electricity infrastructure

**Fig.1.** Impact of i-failures on infrastructures behaviour

**3.1.2 Impact of electricity infrastructure failures (e-failures).** We consider that the occurrence of e-failures leads the electricity infrastructure to be in a partial e-outage state, unless propagation within the infrastructure leads to loosing its control (e.g., a blackout of the power grid), because of an i-failure (this latter case corresponds to escalating events that will be covered in the next section). Fig. 2-a gives the state machine model of the electricity infrastructure taking into account its own failures.

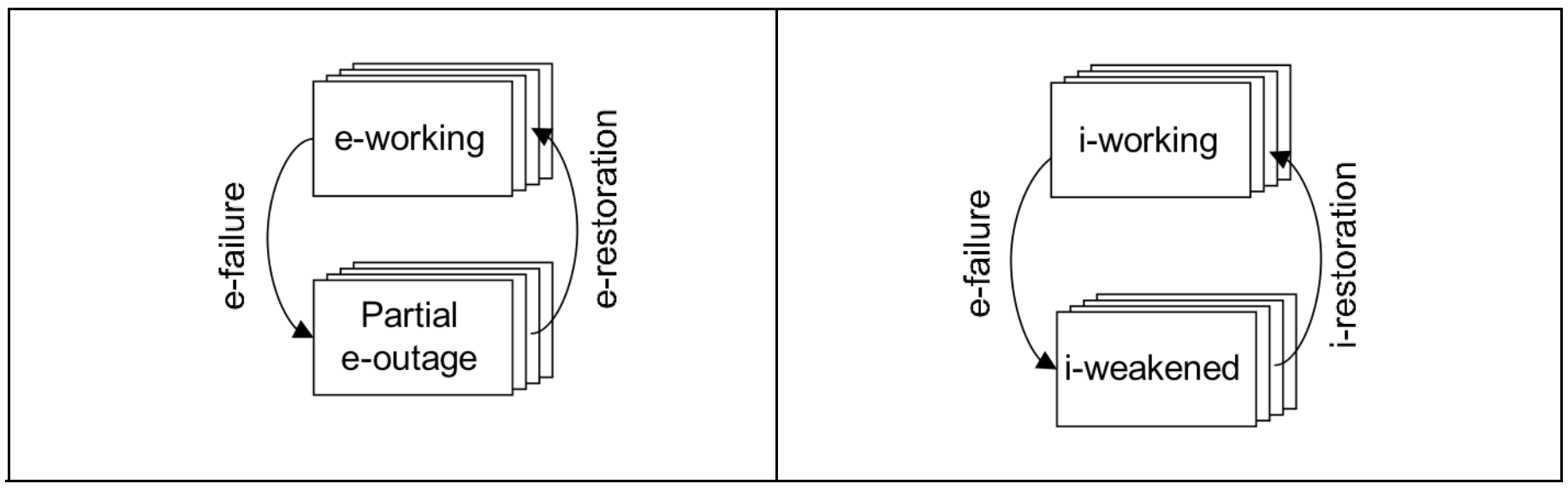

-a- Failure behaviour of the electricity infrastructure

-b- Impact of the electricity infrastructure on the information infrastructure

**Fig. 2.** Impact of e-failures on infrastructures behaviour

Also e-failures may lead the information infrastructure to an i-weakened state in which parts of the information infrastructure can no longer implement their functions, although they are not failed, due to constraints originating from the failure of the electricity infrastructure. Fig. 2-b shows the constraint that the electricity infrastructure puts on the information infrastructure assuming that the latter is in an i-working state.

Tables 2 and 3 summarise the states and events of each infrastructure, taking into account cascading events, as described above.

**Table 2.** States and events of the information infrastructure

| | |
|---|---|
| **Events** | |
| Signalled i-failure | Detected i-failure |
| Masked i-failure | Undetected i-failure |
| i-restotation | Action for bringing back the information infrastructure in its normal functioning after i-failure(s) occurred |
| **States** | |
| i-working | The information infrastructure ensures normal control of the electricity infrastructure |
| Passive latent error | Parts of the information infrastructure have an i-failure, which prevents monitoring of the electricity infrastructure: e-failures may remain unnoticed |
| Active latent error | Parts of the information infrastructure have an i-failure, that may lead to unnecessary, and unnoticed configuration changes |
| Partial i-outage | Parts of the information infrastructure have knowingly an i-failure. Partial i-outage is assumed**:** the variety of functions and of the components of the infrastructure, and its essential character of large network make unlikely total outage |
| i-weakened | Parts of the information infrastructure can no longer implement their functions, although they do not have an e-failure, due to constraints originating from e-failures, e.g., shortage of electricity supply of unprotected parts. |

**Table 3.** States and events of the electricity infrastructure

| | |
|---|---|
| **Events** | |
| e-failure | Malfunctioning of elements of the power grid: production plants, transformers, transmission lines, breakers, etc. |
| e-restoration | Actions for bringing back the electricity infrastructure in its normal functioning after e-failure(s) occurred. Typically, e-restoration is a sequence of configuration change(s), repair(s), configuration restoration(s) |
| e-configuration change | Change of configuration of the power grid that are not immediate consequences of e-failures, e.g., off-line trips of production plants or of transmission lines |
| e-configuration restoration | Act of bringing back the electricity infrastructure in its initial configuration, when configuration changes have taken place |
| **States** | |
| e-working | Electricity production, transmission and distribution are ensured in normal conditions |
| Partial e-outage | Due to e-failure(s), electricity production, transmission and distribution are no longer ensured in normal conditions, they are however somehow ensured, in degraded conditions |
| e-lost | Propagation of e-failures within the electricity infrastructure led to loosing its control, i.e., a blackout occurred. |
| e-weakened | Electricity production, transmission and distribution are no longer ensured in normal conditions, due to i-failure(s) of the information infrastructure that constrain the functioning of the electricity infrastructure, although no e-failure occurred in the latter. The capability of the electricity infrastructure is degraded: lower performance, configuration changes, possible manual control, etc. |

### 3.2. Modelling cascading and escalating failures

The global state machine model of the two infrastructures is built progressively:

- Considering, in a first step, only the constraints of the information infrastructure on the electricity infrastructure.
- Considering constraints of each infrastructure on the other.

Fig. 3 gives a state machine model of the infrastructures, taking into account, only the constraints of the information infrastructure on the electricity infrastructure. The states are described in terms of the statuses of both infrastructures. Both cascading failures (states 3, 4) and escalating ones are evidenced, with a distinction of consequences of the latter in terms of time to restoration (state 6) and of severity (state 7). Dependency of the electricity infrastructure upon the information infrastructure is illustrated by the need for both i- and e-restoration from states 6 and 7.

A noteworthy example of transitions from states 1 to 2, and from 2 to 7 relates to the August 2003 blackout in the USA and Canada: the failure of the monitoring software was one of the immediate causes of the blackout, as it prevented confining the electrical line incident, before its propagation across the power grid [1].

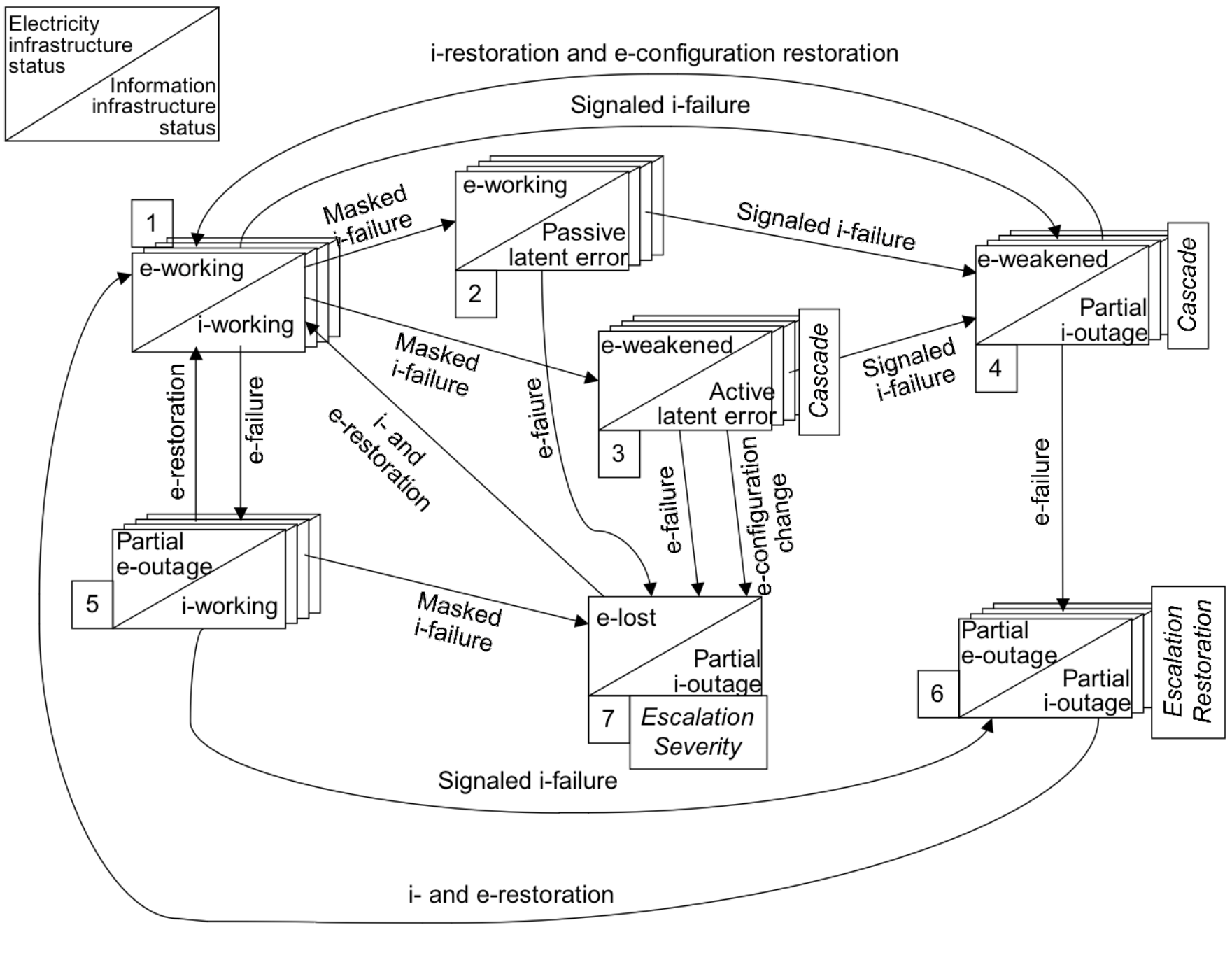

**Fig. 3.** State Machine taking into account constraints of the information infrastructure on the electricity infrastructure

A Petri net representation of the Fig. 3 model is given by Fig. 4 which enables to evidence the cascading and escalating mechanisms. Such mechanisms are, in Petri net terms, synchronizations between the individual events of the infrastructures. Table 4 gives the correspondence between the states and events of Figures 3 and 4.

This Petri net is deliberately kept simple. In particular, it does not distinguish the individual states within a group of states represented by several boxes in Fig. 3. For example, state 2 of Fig. 3 that represents in reality a set of states is represented by a single state in the Petri net of Fig. 4.

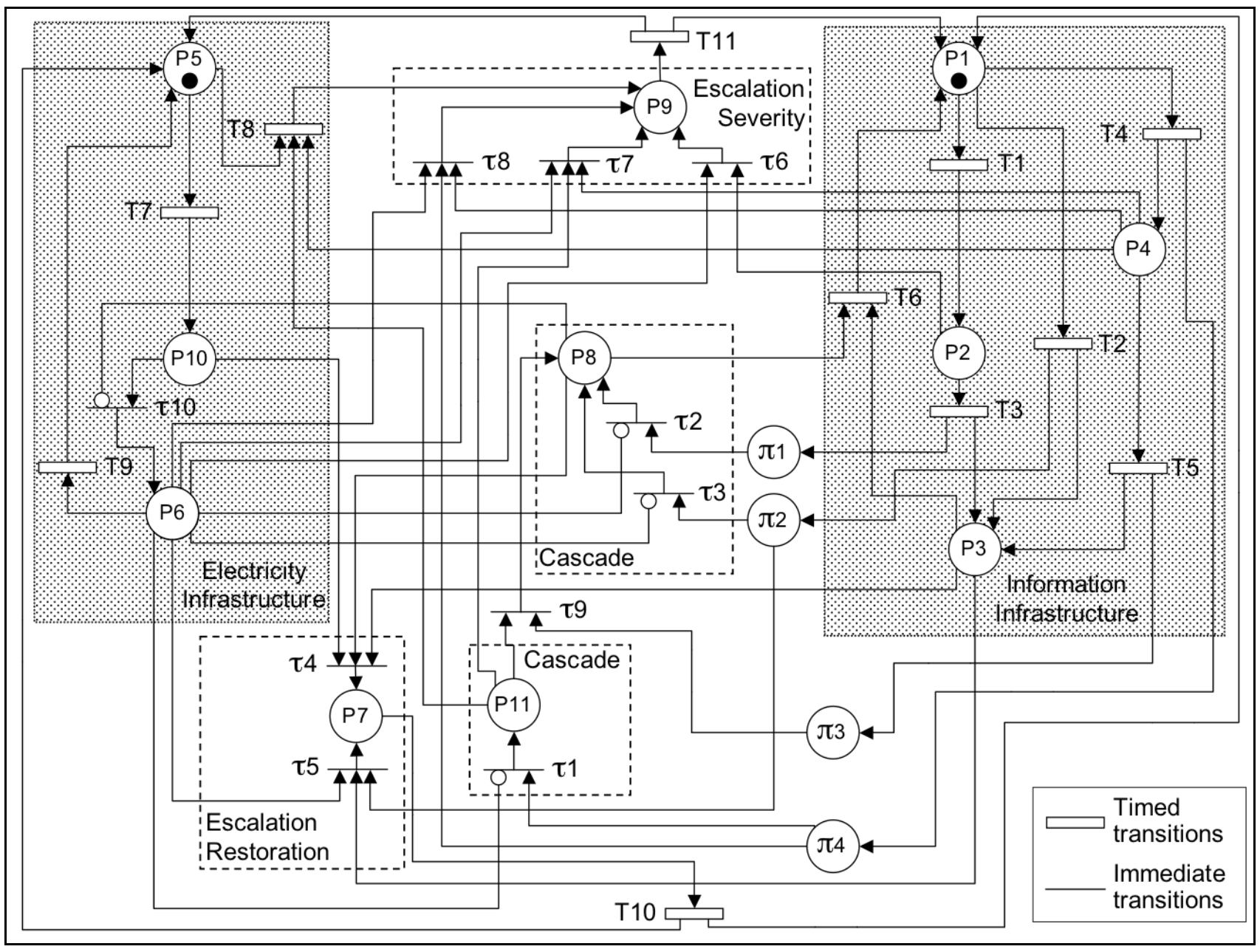

**Fig. 4.** Example of a high level Petri net associated to the model of Figure 3

**Table 4.** Correspondence between states and events of Fig. 3 and Fig. 4

| States in State Machine | Markings in the Petri net |
|---|---|
| 1 | P1, P5 |
| 2 | P2,P5 |
| 3 | P4,P5,P11 |
| 4 | P3,P5,P8 |
| 5 | P1,P6 |
| 6 | P7 |
| 7 | P9 |

| Transitions in State Machine | Transitions in the Petri net |
|---|---|
| 1 → 2 | T1 |
| 1 → 3 | T4 - τ1 |
| 1 → 4 | T2 - τ3 |
| 1 → 5 | T7 - τ10 |
| 2 → 4 | T3 - τ2 |
| 2 → 7 | T7 - τ10 - τ6 |
| 3 → 4 | T5 - τ9 |
| 3 → 7 - configuration change | T8 |
| 3 → 7 - e-failure | T7 - τ10 - τ7 |
| 4 → 1 | T6 |
| 4 → 6 | T7 - τ4 |
| 5 → 1 | T9 |
| 5 → 6 | T2 - τ5 |
| 5 → 7 | T1 - τ6 or T4 - τ8 |
| 6 → 1 | T10 |
| 7 → 1 | T11 |

Fig. 5 gives a state machine model of the infrastructures, taking into account the constraints of the electricity infrastructure on the information infrastructure in addi-tion to those of the information infrastructure on the electricity infrastructure already considered in Fig. 3. In addition, Fig. 5 assumes possible accumulation of e-failures from states 5 to 7 and from the escalation restoration state 6 to the escalation severity state 8.

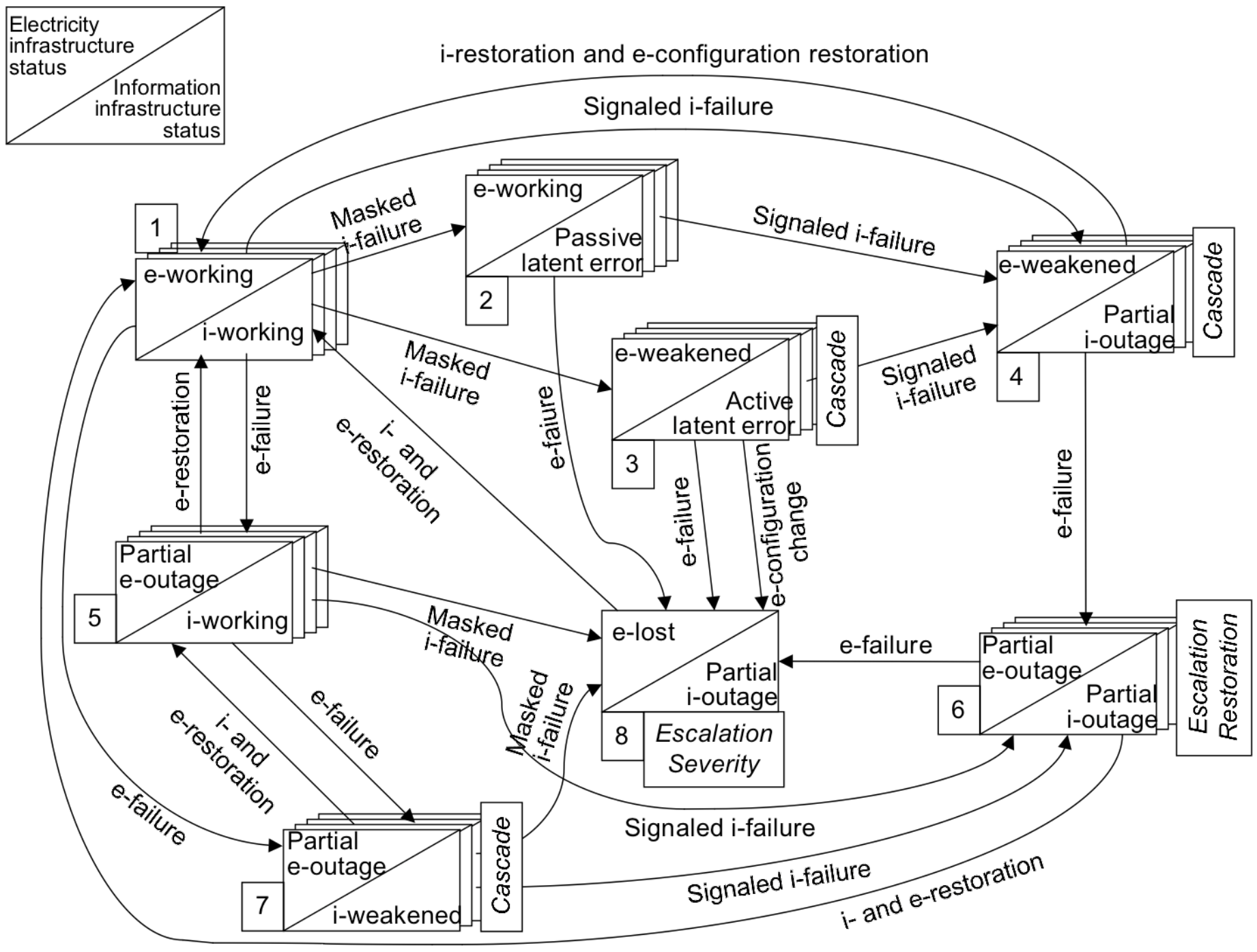

**Fig. 5.** Model of the two infrastructures when considering accidental failures

### 3.3. Modelling common-cause failures

Figure 6 gives a model with respect to common-cause failures that would occur when the infrastructures are in normal operation, bringing the infrastructures into states 6 or 8 of Figure 5, i.e., to escalation. Should such failures occur in other states of the infrastructures of Figure 5 model, they would also lead to states 6 or 8.

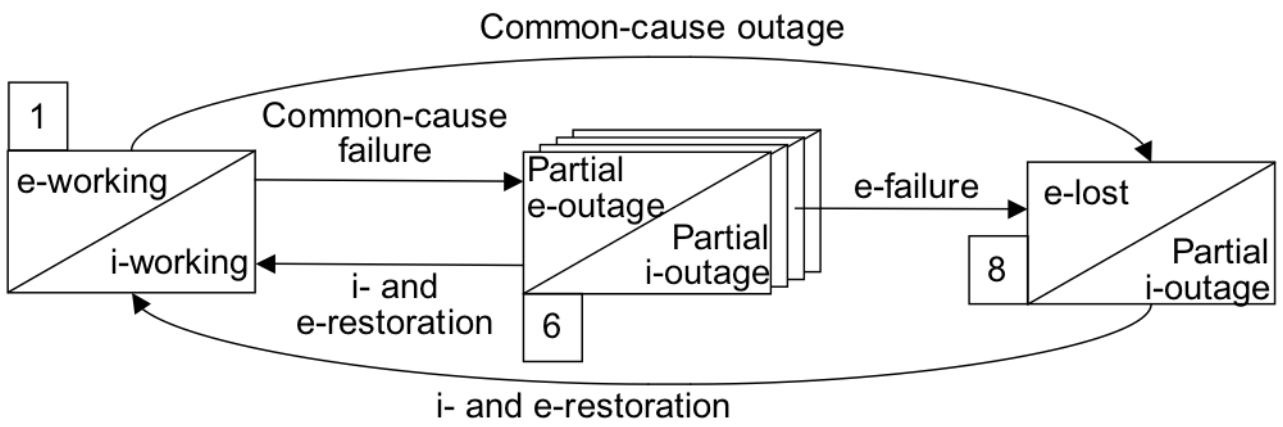

**Fig. 6.** Common-cause failures model

Considering common-cause failures does not introduce additional states, they however add direct transitions from already existing states that do not exist when considering only cascading and escalating failures. The states of the resulting model become almost totally interconnected.

## 4. Malicious attacks of the information infrastructure

We consider malicious attacks of the information infrastructure and their consequences on the electricity infrastructure. Due to the very nature of attacks, a distinction has to be performed for both infrastructures between their real status and their apparent status. For the electricity infrastructure, the apparent status is as reported by the information infrastructure.

Attacks fall into two classes:

- Deceptive attacks that are provoking unperceived malfunctions, thus similar to the latent errors previously considered,
- Percep*tible attacks* creating detected damages.

Deceptive attacks can be:

- *Passive* (i.e., without any direct action on the electricity infrastructure).
- *Active*, provoking configuration changes in the electricity infrastructure.

Fig. 7 gives the state machine model of the infrastructures. This model and the previous one are syntactically identical: they differ by the semantics of the states and of the inter-state transitions. Let us consider for example states 2 and 3.

In state 2, the effects of the passive deceptive attack are: i) the information infrastructure looks like working while it is in a partial i-outage state due to the attack, ii) it informs wrongly the operator that the electricity infrastructure is in partial i-outage, and as consequence iii) the operator performs some configuration changes in the electricity infrastructure leading it to a i-weakened state. Accumulation of configuration changes by the operator may lead the electricity infrastructure into e-lost state.

In state 3, the effects of the active deceptive attack are: i) the information infrastructure looks like working while it is in a partial i-outage state due to the attack, ii) it performs some configuration changes in the electricity infrastructure leading it to a weakened state without informing the operator that the electricity infrastructure is in partial e-outage, for whom the electricity infrastructure appears if it were working. Accumulation of configuration changes by the information infrastructure may lead the electricity infrastructure into a e-lost state.

The difference between states 2 and 3 is that in state 2 the operator has made some actions on the electricity infrastructure and is aware of the e-weakened state, while in state 3 the operator is not aware of the actions performed by the information infrastructure on the electricity infrastructure.

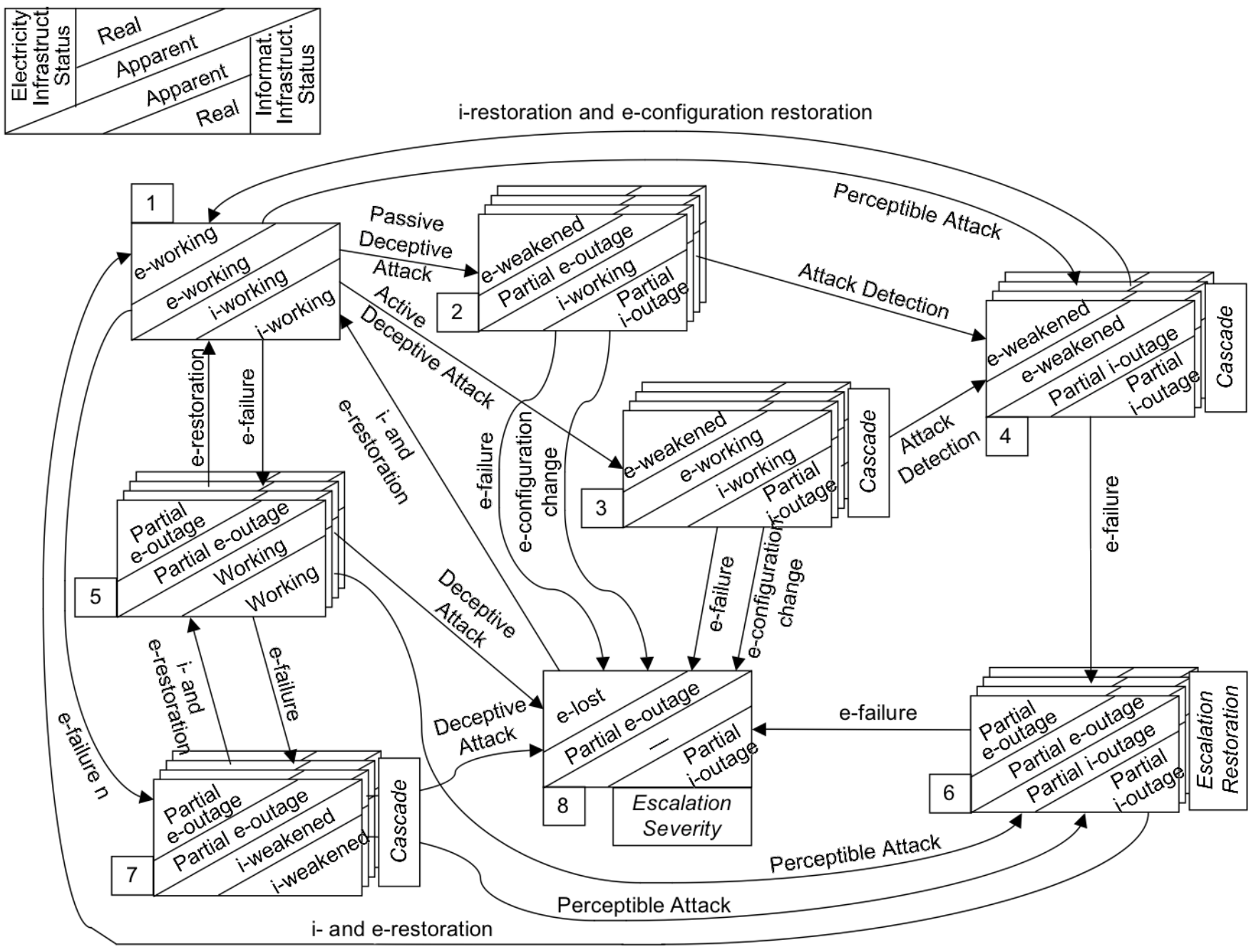

**Fig. 7.** Model of the two infrastructures when considering malicious attacks

After detection of the attack, the apparent states of the infrastructures become identical to the real ones (state 4), in which i-restoration and configuration restoration are necessary to bring back the infrastructures to their working states.

States 5, 6 and 7 are very similar respectively to states 5, 6 and 7 of Fig. 5, except that in state 6 the information infrastructure is in a partial i-outage state following a perceptive attack in Fig. 7 and following a signalled i-failure in Fig. 5.

In Fig. 7, state 8 corresponds to e-lost state but the operator is not aware, he/she has been informed wrongly by the partial i-outage of information infrastructure that it is in a partial e-outage state.

# Conclusion

In this paper we have introduced qualitative models allowing the description and the analysis of the behaviour of the information and electricity infrastructures taking into account the effect of failures of each infrastructure on the other one. These models describe, at a high level, scenarios that may occur when failures occur and the relationship between the states of the two infrastructures.

We have presented different models when considering accidental failures in the information infrastructure and when accounting for malicious attacks. Currently we are investigating a unified model for taking into account both classes of failures of the information infrastructure.

The high level models developed in this paper are to be refined to evaluate quantitative measures characterizing the impact of the interdependency-related failures on the resilience of the electricity and information infrastructures, with respect to the occurrence of critical outages and blackouts. In particular, in parallel to this work, within the CRUTIAL project, preliminary investigations have been carried out in order to develop a hierarchical modelling framework aimed at the detailed modelling of the electricity and information infrastructures taking into account their internal structure, and accidental failures [28]. When considering accidental failures, modelling techniques such as stochastic Petri nets or stochastic activity networks, complemented by data collected from observation of the two infrastructures (or based on more general data, see e.g., [29]) are used. Consideration of malicious attacks raises some difficulties and challenges. Indeed, the very definition of security measures and evaluation has been, and is still, a research topic (see e.g., [30, 31]). Future work will be focussed on the definition of an integrated modelling approach that is well suited to take into account the combined impact of accidental as well as malicious faults.

**Acknowledgement.** This work is partially supported by the European project CRUTIAL (Critical Utility InfrastructurAL Resilience), IST-FP6-STREP - 027513